\documentclass[11pt]{article}
\usepackage{amsmath}
\usepackage{amssymb}
\usepackage{graphicx}

 \advance \textwidth by -2in
 \advance \textheight by -3in
\def\##1{\underline #1}
\def\=#1{\underline{\underline #1}}

\def\eps{\epsilon}
\def\epso{\epsilon_0}

\def\ko{k_0}
\def\lambdao{\lambda_0}

\def\.{\mbox{ \tiny{$^\bullet$} }}

\def\epsa{\epsilon_a}
\def\epsb{\epsilon_b}
\def\epsc{\epsilon_c}

\def\epsdt{\tilde{\epsilon}_d}

\def\ux{\#{u}_x}
\def\uy{\#{u}_y}
\def\uz{\#{u}_z}
\def\up{\#{u}_+}
\def\um{\#{u}_-}

\def\aal{a_L}
\def\aar{a_R}
\def\bbl{r_L}
\def\bbr{r_R}
\def\ccl{t_L}
\def\ccr{t_R}

\def\le{\left(}
\def\ri{\right)}
\def\les{\left[}
\def\ris{\right]}

\def\c#1{\cite{#1}}

\def\r#1{(\ref{#1})}

\begin{document}

\noindent {\bf Stepwise Chirping of Chiral Sculptured Thin Films\\
for Bragg Bandwidth Enhancement}

\vskip 0.4cm

\noindent  { \bf Akhlesh Lakhtakia} 
\vskip 0.2cm
\noindent { Computational \& Theoretical Materials Sciences Group (CATMAS)\\
Department of Engineering Science \& Mechanics\\
212 Earth--Engineering Sciences Building\\
Pennsylvania State University, University Park, PA 16802--6812, USA}

\vskip 0.4cm

\noindent {\bf ABSTRACT:}
{\it A stepwise chirping of the periodicity of a chiral sculptured thin film
is shown to considerably enhance the bandwidth of the Bragg regime,
thereby extending the frequency range of operation as a circular--polarization
filter.}\\

\vskip 0.2cm


\vskip 0.2cm

\section{Introduction}
Theoretical as well as experimental studies have shown that
chiral sculptured thin films (STFs) display the so--called Bragg
phenomenon on axial excitation, 
in consequence of their 
periodically and unidirectionally nonhomogeneous constitution
\c{VLbook, WHL, HWKLR}. Let the 
direction of nonhomogeneity of a chiral STF be parallel to the $z$ axis, while
the film completely occupies the region $0 \leq z \leq L$. When circularly 
polarized, mono\-chromatic light falls normally on this film of sufficient 
thickness, then it is
(i) almost perfectly reflected if the handedness of the incident
light coincides with the structural handedness of the film, and
(ii) almost perfectly transmitted if otherwise~---~provided 
absorption within the film is negligible
and the free--space
wavelength $\lambdao$ of the incident light lies within the so--called Bragg
regime.  Optical devices such as circular--polarization filters and spectral--hole filters 
exploiting the
circular Bragg phenomenon have been fabricated using STF technology,
delivering performance
comparable to commercially available non--STF analogs \c{WHL,HWLM,HWTLM}.

The bandwidth of the
Bragg regime is directly proportional to the local birefringence of the
chiral STF. Although the serial bideposition technique
\c{HWKLR} yields highly birefringent chiral STFs, further enhancement
of the bandwidth can be desirable for certain applications. The objective of this communication
is to show that stepwise chirping of the periodicity of a chiral STF
can easily deliver a considerable enhancement of the bandwidth.

The stepwise chirped chiral STF is described in Section 2, and its
response to  normally incident plane waves is presented in Section 3.
Comparison is made with the response of an unchirped chiral STF.
An $\exp(-i\omega t)$ time--dependence
is implicit in this communication, with $\omega$ as the
angular frequency; vectors are underlined and dyadics are double--underlined; $\#r=x\ux+y\uy+z\uz$ is the
position vector with $\ux$, $\uy$ and $\uz$ as the cartesian
unit vectors; and  $\lambdao$ is the free--space wavelength.

\section{Constitutive Relations}
The nonhomogeneous permittivity
dyadic of a chiral STF is expressed as follows \c{VLbook}:
\begin{equation}
\label{epsbasic}
\=\eps(\#r) = \epso\, \=S_z(z,h)\.\=S_y(\chi)\.
\Big[ \epsa \,\uz\uz +\epsb\,\ux\ux
\,  +\,\epsc\,\uy\uy\Big]\.\=S_y^{-1}(\chi)\.
\=S_z^{-1}(z,h)\, , \quad 0 \leq z \leq L\, .
\end{equation}
In this equation, $\eps_{a,b,c}$ are three relative permittivity
scalars; $\epso = 8.854\times 10^{-12}$~F~m$^{-1}$
is the free--space permittivity; the  tilt dyadic
\begin{equation}
\=S_y(\chi) = \uy\uy + (\ux\ux + \uz\uz) \, \cos\chi + (\uz\ux-\ux\uz)\,
\sin\chi
\end{equation}
represents the locally columnar microstructure of any chiral STF
with $\chi > 0^\circ$; while the rotation dyadic
\begin{equation}
\=S_z(z,h)=
\uz\uz + \le \ux\ux+\uy\uy\ri\,\cos\le\frac{\pi z}{\Omega}\ri \,
+ h\, \le \uy\ux-\ux\uy\ri\,\sin\le \frac{\pi z}{\Omega}\ri
\end{equation}
contains  $2\Omega$ as the structural period, with $h=1$ for
structural right--handedness and $h=-1$ for structural left--handedness.
Taking a whole number of periods to encompass the film thickness,
we choose the ratio $L/2\Omega$ to be a non--prime integer for filtering applications.

The stepwise chirped chiral STF comprises $N_p$ steps, each step having
$\ell_p$ complete periods, with $N_p \ell_p = L/2\Omega$. 
The half--period of the n$^{th}$ step 
is denoted by
\begin{equation}
\Omega_n = \Omega + \frac{2n-N_p-1}{2}\, \delta_\Omega \,,
\quad n \in [1,\,N_p]\,,
\end{equation}
wherein the chirping step $\delta_\Omega$ is selected to be a small fraction
of $\Omega$ to assure that  $\Omega_1 > 0$.
Thus, the expression of the
permittivity dyadic of the stepwise chirped chiral STF is the
same as \r{epsbasic}, except that the dyadic $\=S_z(z,h)$ has to be replaced
by $\=S_z^{chirp}(z,h)$, where
\begin{eqnarray}
\=S_z^{chirp}(z,h)
&=&
\uz\uz + \le \ux\ux+\uy\uy\ri\,\cos\les\frac{\pi (z-\zeta_n)}{\Omega_n}\ris \,
\nonumber
\\
&+& h\,\le \uy\ux-\ux\uy\ri\,\sin\les \frac{\pi (z-\zeta_n)}{\Omega_n}\ris\,,
\quad \zeta_n \leq z \leq \zeta_{n+1}\,,\quad n \in [1,\,N_p]\,,
\end{eqnarray}
and
\begin{equation}
\zeta_n = 2\ell_p (n-1)\le \Omega +\frac{n-N_p-1}{2}\,\delta_\Omega\ri\,
\end{equation}
is the total thickness of the first $n-1$ steps.
Note that the thickness $\zeta_{{N_p}+1}$ of the chirped chiral STF equals
$L$.

\section{Numerical Results and Discussion}
Let an arbitrarily polarized plane wave be normally incident
from the lower half--space $z \leq 0$  on either
of the two films described in the previous section. As a
result, a plane wave is reflected into the lower half--space and another
is transmitted into the upper half--space $z \geq L$.
he electric field phasors associated with the two plane
waves in the lower half--space are stated as  
\begin{equation}
\#{E}_{inc}(\#{r},\lambdao) =
 \le  \aal \, \up + \aar \, \um \ri \,
\exp\le i \ko z \ri  \, ;  \,\, z \leq 0\, ,
\end{equation}
\begin{equation}
\#{E}_{ref}(\#{r},\lambdao) =
\le  \bbl \, \um + \bbr \, \up \ri \,
\exp\le -i \ko z \ri \, ;  \,\, z \leq 0\, ,
\end{equation}
where $\#u_\pm = (\ux \pm i \uy)/\sqrt{2}$ and
$\ko = 2\pi/\lambdao$ is the
free--space wavenumber. Likewise,
the electric field phasor in the upper half--space is
represented as
\begin{equation}
\#{E}_{trs}(\#{r},\lambdao) =
 \le  \ccl \, \up + \ccr \, \um \ri \,
\exp\les i \ko (z -L)\ris  \, ;  \,\, z \geq L \, .
\end{equation}
Here, $\aal$ and $\aar$ are the known amplitudes
of the left-- and the right--circularly polarized (LCP \& RCP)
components
of the incident plane wave;
$\bbl$ and $\bbr$ are the unknown amplitudes
of the reflected plane wave components; while $\ccl$ and $\ccr$
are the unknown amplitudes
of the transmitted plane wave components. 

The reflection amplitudes $r_{L,R}$ and the transmission
amplitudes $t_{L,R}$  can be computed for specified incident amplitudes
($\aal$ and $\aar$) by following standard procedures
described elsewhere; see, e.g., \c{VLbook}. 
Reflection and transmission coefficients are thus defined as the entries
in the 2$\times$2 matrixes in the following two relations:
\begin{eqnarray}
\label{eq15}
\les \begin{array}{cccc} \bbl \\ \bbr  \end{array}\ris  &=&
\les \begin{array}{cccc} r_{LL} & r_{LR} \\ r_{RL} & r_{RR}\end{array}\ris \,
\les \begin{array}{cccc} \aal \\ \aar  \end{array}\ris \, , \\
\label{eq16}
\les \begin{array}{cccc} \ccl \\ \ccr  \end{array}\ris  &=&
\les \begin{array}{cccc} t_{LL} & t_{LR} \\ t_{RL} & t_{RR}\end{array}\ris \,
\les \begin{array}{cccc} \aal \\ \aar  \end{array}\ris
\, .
\end{eqnarray}
The co--polarized transmission coefficients are denoted by $t_{LL}$ and
$t_{RR}$,
and the cross--polarized ones by $t_{LR}$ and $t_{RL}$; and similarly for the
reflection coefficients in \r{eq15}.

The reflectances $R_{LL} = \vert r_{LL}\vert^2$, etc., and the
transmittances $T_{LL} = \vert t_{LL}\vert^2$, etc., were calculated
as functions of the free--space wavelength $\lambdao$ for a chiral STF
described by the following parameters: $h=1$, $\eps_a = 2.8$, $\eps_b = 3.4$,
$\eps_c = 2.9$, $\chi = 20^\circ$, $\Omega = 155$~nm, and $L = 126 \,\Omega$.
Graphs are presented in Figure 1, which clearly show the Bragg regime 
delineated by $\lambdao \in [528,\,564]$~nm almost agrees
with the theoretical range $\lambdao \in \les 2\Omega\epsc^{1/2},\,
2\Omega\epsdt^{1/2}\ris$ \c{VLbook}.
In this regime,
$R_{RR}$ and $T_{LL}$ are very high, while $R_{LL}$ and $T_{RR}$
are very low, because the chosen film is structurally right--handed.

For the stepwise chirped chiral STF, we chose $N_p=21$, $\ell_p = 3$
and $\delta_\Omega = 0.5$~nm. Thus, the smallest half--period $\Omega_1=150$~nm
and the largest half--period $\Omega_2=160$~nm, while the average
half--period is 155~nm (the same as $\Omega$ for Figure 1). 
The computed reflectances
and transmittances, plotted in Figure 2 as functions of $\lambdao$, show 
that the Bragg regime now
is $\lambdao \in [514,\,580]$~nm.  The lower limit of the Bragg regime
is somewhat higher than $2\Omega_1\epsc^{1/2}$, and the upper limit is
somewhat lower than $2\Omega_{N_p}\epsdt^{1/2}$.

Relative
to the 36~nm bandwidth in Figure 1, the Bragg regime in Figure 2
has a bandwidth of 66~nm~---~an enhancement by more than 80\%.
The enhancement comes without any noticeable reduction in the
polarization--sensitivity of the film. Furthermore, calculations show
that mild dispersion and absorption do not affect the enhancement
significantly.

Stepwise changes in the periodicity of a chiral STF are likely to be
accompanied by changes in the relative permittivity scalars $\eps_{a,b,c}$,
which attribute of STFs was not accommodated in the theoretical study presented.
Because the chirping step is very small compared with the mean half--period
$\le \Sigma_{n=0}^{N_p}\,\Omega_n\ri /N$ however, those changes are expected
have insignificant effect. 

To conclude, the presented theoretical study has demonstrated that
a stepwise chirping of the periodicity of a chiral sculptured thin film
considerably enhances the bandwidth of the Bragg regime,
thereby extending the frequency range of operation as a circular--polarization
filter.

\newpage

\begin{figure}[!htb]
\centering
 \includegraphics[width=12.5cm]{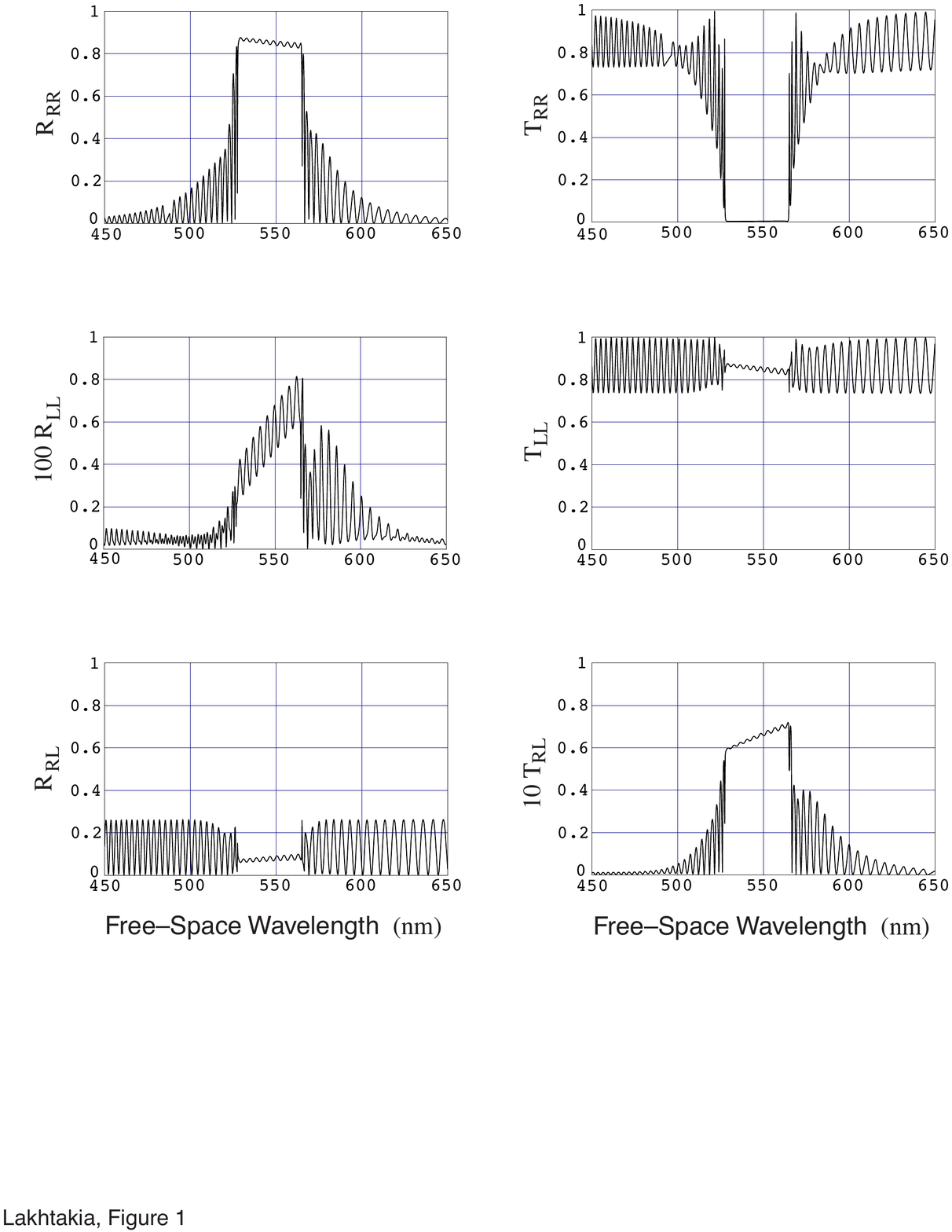} 
 \caption{Computed spectrums of the reflectances $R_{RR}$, $R_{LL}$
and $R_{LR}=R_{RL}$ and the transmittances $T_{RR}$, $T_{LL}$
and $T_{LR}=T_{RL}$ of a chiral STF described by the following
parameters: $h=1$, $\eps_a = 2.8$, $\eps_b = 3.4$,
$\eps_c = 2.9$, $\chi = 20^\circ$, $\Omega = 155$~nm, and $L = 126 \,\Omega$.
    }
\end{figure}

\begin{figure}[!htb]
\centering
 \includegraphics[width=12.5cm]{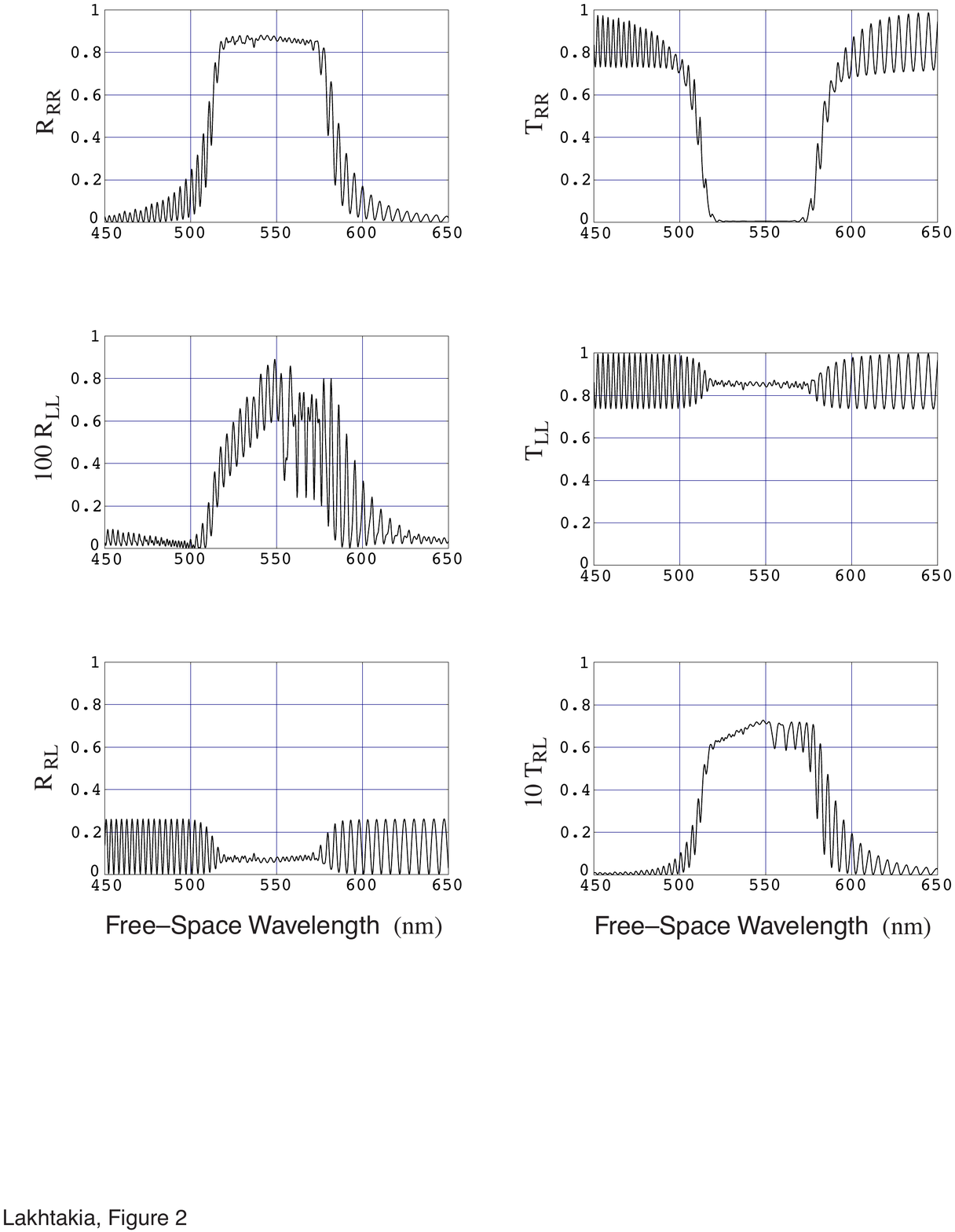} 
 \caption{Same as Figure 1, except for a stepwise chirped chiral STF
described by the following parameters: $h=1$, $\eps_a = 2.8$, $\eps_b = 3.4$,
$\eps_c = 2.9$, $\chi = 20^\circ$, $N_p = 21$, $\ell_p = 3$, $\delta_\Omega=0.5$~nm, and $L = 126 \,\Omega$.
    }
\end{figure}

\end{document}